\documentclass{article}

\usepackage[preprint,nonatbib]{neurips_2025}
\usepackage[numbers]{natbib}

\usepackage{enumitem}
\usepackage[utf8]{inputenc} 
\usepackage[T1]{fontenc}    
\usepackage{hyperref}       
\usepackage{url}            
\usepackage{booktabs}       
\usepackage{amsfonts}       
\usepackage{amsthm}
\usepackage{nicefrac}       
\usepackage{microtype}      
\usepackage{xcolor}         
\usepackage{algorithm}
\usepackage{algpseudocode}
\usepackage{amsmath}
\usepackage{caption}
\usepackage{graphicx}
\usepackage{tabularx}

 \usepackage{inconsolata}
 \usepackage{comment}

\usepackage[usenames,dvipsnames,svgnames,x11names]{xcolor}
\usepackage{amssymb}

\colorlet{RED}{red}
 
\definecolor{trcolor}{rgb}{0.7,0.3,0.7}
\definecolor{shanks}{rgb}{0.7,0.2,0.4}
\definecolor{dgcolor}{rgb}{0.5,0.3,0.7}
\definecolor{nccolor}{rgb}{0,0,1}
\definecolor{gacolor}{rgb}{0.5,0.6,1}
\definecolor{sscolor}{rgb}{0.4,0.4,0.5}

\title{Position: Enterprise AI Must Enforce Participant-Aware Access Control}

\author{%
Shashank Shreedhar Bhatt \quad Tanmay Rajore \quad Khushboo Aggarwal \And Ganesh Ananthanarayanan \quad Ranveer Chandra \quad Nishanth Chandran \And Suyash Choudhury \quad Divya Gupta \quad Emre Kiciman \And
Sumit Kumar Pandey \quad Srinath Setty \quad Rahul Sharma \quad Teijia Zhao \\\\
Microsoft Corporation \\\\
\texttt{\{t-shabhatt, t-trajore, khaggarw, ga, ranveer, nichandr, suchoudh,}\\
\texttt{digup, emrek, sumpandey, srinath, rahsha, tiejiazhao\}@microsoft.com}
}

\newtheorem{theorem}{Theorem}

\begin{document}

\maketitle

\begin{abstract}
  Large language models (LLMs) are increasingly deployed in enterprise settings where they interact with multiple users and are trained or fine-tuned on sensitive internal data. While fine-tuning enhances performance by internalizing domain knowledge, it also introduces a critical security risk: leakage of confidential training data to unauthorized users. These risks are exacerbated when LLMs are combined with Retrieval-Augmented Generation (RAG) pipelines that dynamically fetch contextual documents at inference time.

  We demonstrate data exfiltration attacks on AI assistants where adversaries can exploit current fine-tuning and RAG architectures to leak sensitive information by leveraging the lack of access control enforcement. We show that existing defenses, including prompt sanitization, output filtering, system isolation, and training-level privacy mechanisms, are fundamentally probabilistic and fail to offer robust protection against such attacks.

{\bf We take the position that only a deterministic and rigorous enforcement of fine-grained access control during both fine-tuning and RAG-based inference can reliably prevent the leakage of sensitive data to unauthorized recipients.}

%
We introduce a framework centered on the principle that any content used in training, retrieval, or generation by an LLM is explicitly authorized for \emph{all users involved in the interaction}. Our approach offers a simple yet powerful paradigm shift for building secure multi-user LLM systems that are grounded in classical access control but adapted to the unique challenges of modern AI workflows. Our solution has been deployed in Microsoft Copilot Tuning, a product offering that enables organizations to fine-tune models using their own enterprise-specific data.

\end{abstract}

\section{Introduction}

Large language models (LLMs) are increasingly being integrated into enterprise workflows to support tasks such as code completion, document generation, and email drafting—e.g. via modern AI productivity tools like Cursor AI~\cite{cursor}, Microsoft Copilot~\cite{MicrosoftCopilot}, Google Gemini Advanced~\cite{GoogleWorkspaceAI}, etc. These tools are frequently deployed in enterprise settings with multi-user environments, where data is protected by strict access control mechanisms tied to individual user permissions. To enhance the effectiveness of these AI tools, enterprises often employ models that are trained or fine-tuned on internal data, which varies in sensitivity and is typically siloed across departments or users. While fine-tuning incorporates valuable domain-specific knowledge, it also introduces a critical risk: the potential leakage of confidential information to unauthorized users.

Models fine-tuned on sensitive information—such as HR records, internal policies or high-impact business strategies—may inadvertently reveal this information when users interact with the AI tools and query the models, particularly in environments with heterogeneous access privileges. Worse still, such leakage can occur not only through targeted extraction attacks but also unintentionally during routine interactions, like drafting emails or querying an assistant on personal and shared documents~\cite{leakagenews}. Feedback from numerous business groups highlights that the risk of data leakage and the absence of robust security controls remain significant barriers to enterprise adoption of AI tools~\cite{register}.

These risks are further amplified when fine-tuned models are integrated with Retrieval Augmented Generation (RAG) pipelines. In such systems, user prompts are augmented with documents retrieved based on query context, improving the response quality. 
However, if the retrieval mechanism fails to strictly enforce access control lists (ACLs) for \emph{all users involved in the interaction}, it may introduce restricted content into the model’s context, leading to unauthorized disclosures. For instance, consider a scenario where Alice uses a RAG-based tool to draft a reply to Bob's email. If ACLs for both Alice and Bob are not properly enforced, a malicious Bob can exploit the retrieval process to exfiltrate sensitive data—embedding hidden instructions in his email and then decoding the model’s response via steganographic techniques. Alarmingly, such attacks can succeed even when Bob's incoming email and assistant's outputs are subject to human review (\S~\ref{sec:xpia}).

\textbf{We take the position that deterministic and provable enforcement of fine-grained access controls is essential for the safe deployment of LLMs in multi-user environments such as enterprises. 
This stands in direct contrast to probabilistic defenses---such as input sanitization, output filtering, system-level isolation, and training-level techniques---which, while useful, are fundamentally insufficient due to their lack of rigorous end-to-end guarantees.
We advocate a simple and robust principle: RAG pipelines and fine-tuned models must enforce access control {\em at every stage of processing}, ensuring that any content incorporated into a model’s output is authorized for access by \emph{all intended recipients}. }

By reframing data security for LLMs as a problem of principled access control, we introduce a new design paradigm—one that prioritizes auditability and provable security guarantees. We also present a framework that demonstrates how access control can be enforced both during fine-tuning and at inference time within RAG pipelines.

\textbf{Overview.}
First, we show that, in the absence of deterministic safeguards on fine-tuning data and fine-tuned model access, fine-tuned models are vulnerable to data extraction attacks—risks that existing defenses fail to mitigate (\S \ref{sec:finetuning-attacks}). We then introduce access control-aware fine-tuned models (\S \ref{sec:access-FT}), where the fine-tuning process inherits the permissions and ACLs of the underlying training data. Specifically, we enforce that a user gets access to a fine-tuned model if and only if they have access to {\em all} data used during fine-tuning. Na\"{\i}vely, if all private data within an organization, that is subject to fine-grained access controls, is used for fine-tuning, it may result in no single user (or very few users) being authorized to access the resulting model. To address this, we formulate the joint selection of training data and authorized model users as the problem of finding bicliques in a bipartite graph. Our initial solution has been deployed in Microsoft's Copilot Tuning~\cite{copilottuning}, which enables customers to fine-tune models on their sensitive data. Finally, we study information leakage in LLM-assisted systems during inference in the presence of multiple users, and discuss how enforcing deterministic access checks provides a principled solution to eliminate information leakage (\S \ref{sec:RAG}). Related work on data leakage attacks, existing defenses, and their limitations is discussed inline within the relevant sections and scenarios, rather than in a standalone section.


\section{Data extraction attacks on fine-tuned models}
\label{sec:finetuning-attacks}
Fine-tuning is the process of taking a pre-trained machine learning model—such as a large language model (LLM)—and continuing its training on a smaller, task-specific dataset~\cite{treviso2023efficient}. This allows the model to adapt to the nuances and requirements of a particular application or domain, improving its performance in that context.
Fine-tuning allows enterprises to leverage their private, often sensitive data to create models tailored to their needs—for example, code completion using internal repositories and document generation following company-specific templates.
While techniques like Retrieval-Augmented Generation (RAG) can provide application-specific context at inference time, fine-tuning has been shown to more effectively embed domain knowledge into a model~\cite{balaguer2024rag}. RAG is better suited as a complementary approach to support fine-tuned models rather than replace them. Additionally, fine-tuning can reduce the context needed during inference, which lowers inference costs~\cite{zou2024promptintern}.


While offering substantial value in enterprise settings, fine-tuning introduces a unique security risk. Since employees often have varying levels of access to sensitive data, a fine-tuned model could unintentionally expose restricted information. This makes it a potential attack vector for data exfiltration, as models are known to leak training data through their inference outputs~\cite{carlini2021extracting, carlini2023usenix, nasr2023scalable,SongVLDB}.


\noindent\textbf{Internal data leakage via fine-tuned models.} Consider the following scenario: An enterprise fine-tunes a model $M$ on its internal data, including sensitive information $D$ that employee Bob is not authorized to access. 
If Bob is subsequently granted access to the fine-tuned model $M$, he may, either inadvertently or deliberately, extract privileged information contained in $D$ through model interactions, resulting in a breach of internal data confidentiality or data leakage.
For example, suppose $D$ includes comprehensive HR data, such as compensation policies, and Bob is a non-HR employee without access privileges to this subset of information. If $M$ is fine-tuned on $D$, it may implicitly encode and potentially reveal these sensitive HR policies during inference to Bob, thus circumventing established access controls.

This form of leakage arises from the very objective of fine-tuning: to enhance model performance by internalizing attributes of the training distribution. While prior work has demonstrated that models can memorize and regurgitate sensitive training data verbatim during inference \cite{carlini2021extracting, carlini2022quantifying, nasr2023scalable, mccoy2023much, inan2021training, tirumala2022memorization, ippolito2022preventing, kiyomaru2024comprehensive, carlini2023usenix, SongVLDB}, we emphasize that data leakage can occur even in the absence of exact memorization. Subtler forms of leakage—such as the paraphrased disclosure of sensitive content—can also arise due to the model’s generalization over restricted training signals.
Importantly, the class of sensitive information under consideration extends beyond personally identifiable information (PII). It also encompasses business-critical data—such as internal policies, strategic plans, or confidential departmental communications—that may be selectively accessible within different organizational units. When a model is fine-tuned on data sourced from multiple such units without access control enforcement, it effectively flattens internal data silos, exposing restricted information to users across the enterprise regardless of their authorization level.


There are a variety of attacks that aim to learn information about data used during training process. These include query-based attacks that can directly extract training data \cite{carlini2021extracting, mccoy2023much, nasr2023scalable, inan2021training, tirumala2022memorization, SongVLDB}, as well as membership inference attacks that aim to determine whether specific data was part of the training set \cite{carlini2021extracting, mireshghallah2022quantifying, fu2023practical, SongVLDB}.
Alarmingly, memorization has been shown to increase with the current LLM scaling trends—both in terms of larger model sizes \cite{carlini2021extracting, carlini2022quantifying, kiyomaru2024comprehensive, tirumala2022memorization, SongVLDB} and larger training datasets \cite{SongVLDB, carlini2022quantifying, kiyomaru2024comprehensive}.
This poses a serious risk in multi-user settings, where a model fine-tuned on data with varying access controls is made available to all users. Crucially, data leakage can occur not only through targeted attacks, but also inadvertently during normal interactions by legitimate users. Therefore, models fine-tuned on sensitive data must be treated as potential sources of data leakage.



\textbf{Inadequacy of current defense mechanisms. }
Despite the aforementioned risks, existing defense mechanisms remain inadequate for preventing the leakage of sensitive fine-tuning data—particularly in enterprise settings, where such data often encapsulates complex, structured information beyond memorizable facts or personally identifiable information.
Defense strategies can be broadly classified by their intervention stage \cite{sousa2023keep}: (i) data-centric approaches that preprocess fine-tuning data to mitigate leakage risk, (ii) output-level methods that apply post hoc sanitization or filtering to model predictions, and (iii) training-level techniques that modify the learning dynamics or loss function to enhance robustness against leakage.

Data-centric approaches aim to mitigate leakage risks by preprocessing or sanitizing training data prior to fine-tuning. N-gram deduplication \cite{lee2021deduplicating, kandpal2022deduplicating} reduces memorization by eliminating repeated sequences, while data scrubbing techniques \cite{lukas2023analyzing, vakili2022downstream} target the removal of personally identifiable information (PII). Obfuscation methods \cite{zhang2018privacy} perturb the input distribution by injecting noise. Although such techniques can reduce the likelihood of memorizing specific data points, they do not offer any formal guarantees. Moreover, in enterprise settings—where fine-tuning data often encodes sensitive, structured information—these methods are insufficient
and frequently inapplicable.


Output-level methods focus on post-processing or controlling model outputs at inference time to mitigate the risk of sensitive information disclosure. Techniques such as confidence masking \cite{jia2019memguard} introduce noise into model outputs to obscure potential memorized content, but often degrade utility without providing robust guarantees against leakage. Data Loss Prevention (DLP) tools \cite{hart2011text, neerbek2018rnn, purview} aim to detect and prevent the exposure of sensitive content—such as PII, protected health information~(PHI), or financial data—through pattern matching or rule-based classification. However, DLP methods are fundamentally limited: they cannot determine whether a model’s response to a particular user (e.g., Bob) is derived from training data that is inaccessible to that user. Crucially, they lack the ability to track fine-grained information flow from training data to specific model outputs.

Approaches that modify the training procedure to mitigate information leakage include regularization and differential privacy (DP).
Regularization encompasses techniques aimed at reducing overfitting, thereby limiting the model’s tendency to memorize individual training examples~\cite{yin2021defending, li2021membership}. However, applying this technique in our setting—where the goal is to prevent the model from learning implicit, large-scale sensitive information across an enterprise—would necessitate overly aggressive regularization, severely compromising the model’s utility for authorized users.
Differential privacy offers a more formal approach to limiting information leakage \cite{dwork2006dp, ramaswamy2020training, kandpal2023user}. A randomized algorithm is said to be differentially private if its output distributions remain similar when applied to two datasets differing in a single data point. While conceptually appealing, DP faces several limitations in the context of fine-tuning large models. First, even a differentially private training algorithm allows bounded leakage, captured by the privacy parameter $\epsilon$, which often must be large to maintain acceptable utility. Second, standard DP protects only individual data samples. In practice, sensitive enterprise data may span correlated groups of records, all of which may be inaccessible to a given user. Protecting such large-scale information requires group differential privacy \cite{dwork2014algorithmic}, which increases the privacy budget $\epsilon$ and degrades utility to the point that the method becomes nonviable for real-world deployment.


\textbf{Summary.} Providing model access in a multi-user setting—where the model has been fine-tuned on sensitive data governed by fine-grained access controls—inevitably carries the risk of unintended data exposure.
Existing defense mechanisms, whether applied at the input, output, or fine-tuning stages, lack end-to-end guarantees and fail to provide reliable protection against data leakage. This fundamental security gap remains a key barrier to the widespread adoption of fine-tuned LLMs in enterprise settings.

\newcommand{\docs}{\tilde{D}}
\newcommand{\users}{\tilde{U}}
\newcommand{\entities}{\tilde{E}}

\newcommand{\docsall}{\mathcal{D}}
\newcommand{\entall}{\mathcal{E}}

\section{Access control aware fine-tuned models}
\label{sec:access-FT}
In \S\ref{sec:secp}, we describe our security principle that provably prevents data leakage from fine-tuned models. Next, in \S\ref{sec:setup}, we discuss how access control lists (ACLs) can be used to realize this principle followed by a graphical representation of the problem that we need to solve.
We then present several example solutions as proofs of concept (PoCs) that illustrate our approach to building useful fine-tuned models in multi-user environments (\S\ref{sec:uses}). One of these has been deployed in the AI assistant of a large enterprise solution
provider for real-world customers. 

\subsection{The security principle}
\label{sec:secp}

\noindent\textbf{Formal security requirement.}
We require that a user $U$'s access to a fine-tuned model does not grant $U$ access to any unauthorized information. This requirement formally ensures that the use of fine-tuned models does not result in data leakage attacks.

Next, we state our security principle that provably satisfies the above security requirement.

\noindent\textbf{Security principle for fine-tuning.} Let $M$ be a model obtained by fine-tuning a publicly pre-trained model on a collection of sensitive documents $\docs$. Then, $M$ can be provided to a user $U$ only if $U$ is authorized to access all documents in $\docs$.

For instance, if $M$ is a model fine-tuned on sensitive data that is accessible to both Alice and Charlie and not accessible to Bob, then $M$ can be provided to Alice and Charlie and not to Bob. 
Conversely, to fine-tune a model intended for a user group $\users$, the training data must be restricted to documents that are accessible to all users $U \in \users$.

Enforcing our security principle when building fine-tuned models clearly requires careful examination of who has access to the underlying training data. To this end, we examine the access control lists (ACLs) associated with data in commonly used file systems, such as SharePoint sites, Google Workspace, and similar platforms. Next, we review a few definitions and provide a graphical representation of our problem.

\subsection{Bicliques are secure for fine-tuning}
\label{sec:setup}
\textbf{Notation and definition.} 
We denote documents by $D$, entities by $E$, users by $U$, and models by $M$. 
An entity $E$ could either be a single user or a group of users (e.g., distribution lists or Active Directory security groups). 
Each document $D$ is associated with an {\em access control list (ACL)} consisting of a list of entities $E$ who are authorized to access $D$.
A {\em bipartite graph} $G$ is a graph in which vertices can be divided into two disjoint sets such that no two vertices in the same set are connected.
A biclique is a complete bipartite graph where each vertex in one set is connected to each vertex in other set.

\textbf{ACLs as a graph.} Given a database of document IDs $\docsall$ with their corresponding ACLs denoting entities from universe $\entall$, we can represent this access control information using a bipartite graph $G$ between $\docsall$ and $\entall$ such that $D\in \docsall$ with ACL $A$ is connected to $E\in\entall$ if and only if $E \in A$. 

The following theorem captures the idea that bicliques can used to select documents for fine-tuning that respects access control.

\begin{theorem} Let $G$ be the bipartite graph over $\docsall \cup \entall$ using the given ACLs and $B$ be a biclique in $G$ with vertices $\docs \subseteq \docsall$ and $\entities \subseteq \entall$. Then, a fine-tuned model $M$ satisfying the security principle in Section~\ref{sec:secp} can be created for the set of entities $\entities$ by fine-tuning on documents in $\docs$.
\end{theorem}

It is straightforward to see the validity of this theorem as by design every entity in $\entities$ is authorized to access every document in $\docs$.

\subsection{Our approach to building useful fine-tuned models for multi-user environments}
\label{sec:uses}

In the previous subsections, we established a security principle for access-control-aware fine-tuning that provably prevents all data leakage scenarios described in Section~\ref{sec:finetuning-attacks}. We also showed that bicliques can be leveraged to enable secure fine-tuning. In this section, we address the question of model utility: given the potentially large number of bicliques in the ACL-derived graph, which fine-tuned models are actually worth building?

At one extreme, we could fine-tune a separate model for each user. While this offers maximum personalization, it is prohibitively expensive and operationally unsustainable for any enterprise. A more practical goal is to fine-tune models that can be shared across many users/entities—a property we refer to as {\em entity coverage.} For example, we could train a single model for an entire enterprise, achieving $100\%$ entity coverage. However, our security model restricts training data to documents that are "public" within the organization. This constraint may exclude a large portion of relevant content, limiting the model's utility for enterprise-specific tasks. To maintain task relevance, we must selectively include documents containing essential domain knowledge-a property we refer to as {\em document coverage}. This, in turn, may limit which users are permitted access to the fine-tuned model, and reduce entity coverage. Thus, there is a tradeoff between entity coverage and document coverage.

We consider two primary use cases where a model maker or an IT administrator aims to fine-tune a model for an enterprise setting. The IT administrator provides a set of document IDs $\docsall$ along with their corresponding access control lists (ACLs) defined over a set of entities $\entall$.  

\textbf{IT administrator specifies a target entity set $\entities$ for model access.}
This case is straightforward: we select all documents $D \in \docsall$ whose ACL $A$ satisfies $\entities \subseteq A$. This approach is efficient, requiring only a linear scan over all document IDs and their associated ACLs.

\textbf{IT administrator does not specify any preference.}
In this scenario, the objective is to identify a subset of documents and entities that achieves a balance between document coverage and entity coverage, as discussed earlier. We propose two candidate strategies to address this:

1) {\em Maximum biclique:} A biclique with the largest number of edges is referred to as a maximum biclique. This configuration is desirable for fine-tuning as it offers broad document and entity coverage. However, identifying a maximum biclique is known to be NP-Complete~\cite{PEETERS2003651}. While algorithms such as branch-and-bound techniques~\cite{mbc,z3} can be employed to find a maximum biclique, they are computationally expensive, especially in dense graphs—a situation commonly encountered in enterprise settings.

2) {\em Maximal biclique:} Intuitively, this approach involves generating candidate maximal bicliques\footnote{Here, bicliques to which no more documents or entities can be added without violating the biclique property are called maximal bicliques.} and selecting the one with the maximum number of edges. As a heuristic, we consider each unique ACL in the graph as a candidate entity set $\entities$, and apply the strategy described in the first case to determine the corresponding set of documents $\docs$. We then attempt to expand the set of entities by including any other entity that has access to all documents in $\docs$, thereby forming a candidate maximal biclique. This heuristic is particularly relevant for enterprise systems where it is common for a few entity sets to govern many documents, increasing its likelihood of identifying the true maximum biclique (see Appendix \ref{appendix} for a detailed discussion).


\textbf{Real-world adoption.} Building on the ideas outlined above for a maximal biclique, our initial solution (described in Appendix \ref{appendix}) is publicly offered as part of Microsoft's Copilot Tuning. This product enables large enterprise customers to train fine-tuned models on their proprietary data.



\section{Eliminating information leakage in multi-user RAG systems}
\label{sec:RAG}
AI assistants are commonly enhanced with Retrieval-Augmented Generation (RAG), which supplements the LLM’s prompt with relevant context. We consider settings in which multiple users interact with AI assistants—providing inputs, receiving outputs, and often influencing shared system state. While the descriptions in this section primarily assume public LLMs for inference after the retrieval by the RAG, we also touch upon the implications of using {\em fine-tuned models along with RAG} in \S\ref{sec:access-RAG}. 
Securing fine-tuned models in conjunction with RAG-based retrieval in multi-user settings is a highly complex challenge that demands rigorous, carefully designed, and deterministically enforced access control mechanisms to prevent unintended data exposure.
%
Hence for ease of exposition, we use public LLMs for our discussion on multi-user RAG-based AI assistants (till we get to the end of \S\ref{sec:access-RAG}).

We consider two representative scenarios: (a) an email assistant that helps a user draft replies to emails received from various senders, and (b) collaborative writing tools where multiple users co-author content using an LLM assistant. We start by focusing on the email assistant scenario, and later briefly outline how similar vulnerabilities manifest in collaborative writing tools.


\subsection{Email Assistant with RAG + LLM}
\label{sec:email}

Consider an LLM-powered email assistant designed to help users draft, edit, and summarize emails. Retrieval-Augmented Generation (RAG) enriches the LLM’s prompt with documents from the user's mailbox or Google Drive as in Google Gemini Advanced. We begin by describing two common modes in email assistants and how RAG operates in each mode.

In the {\em agentic mode}, the assistant acts as a drafting partner. When Alice invokes the system - for example, by saying “Help me draft a reply to Bob” — the assistant generates a candidate message for her review. Alice retains final control: she can edit, discard, or approve the draft. This review step provides a clear checkpoint, ensuring that no content is sent without Alice’s explicit consent\footnote{As we will see later, even when Alice retains final control, an adversarial sender Bob can exfiltrate sensitive information from Alice's inbox.}.

In contrast, the {\em fully agentic mode} removes the manual checkpoint entirely. When Alice issues a high-level directive, the assistant drafts and sends emails directly on her behalf. This maximizes convenience and speed, but places full trust in the assistant to avoid errors or unintended disclosures—even in sensitive contexts.

At the core of both modes is a RAG pipeline. The assistant first indexes Alice’s mailbox—subject lines, message bodies, and attachments—by embedding each item into a semantic vector space. When Alice issues a prompt, the system embeds this as well, and performs a semantic search over the index, retrieving the top-k most relevant items. These retrieved snippets are then prepended to the LLM prompt~\cite{rag_definations}, grounding the model in specific details from past conversations, dates, or figures that enhance response quality.

While RAG can substantially improve assistant performance, it also broadens the attack surface: any document in Alice’s mailbox may be pulled into the LLM’s context if the retrieval mechanism is not carefully constrained. This vulnerability becomes particularly serious when an adversarial sender—say, Bob—injects a malicious prompt into an email, aiming to influence what the system retrieves and how it responds.

We introduce our two principal actors and define the attack class. {\em Alice} is a benign user of the email assistant. She interacts in good faith, trusts the system to help her draft and send emails, and never intends to leak private data. In the agentic mode, she manually reviews every draft; in the fully agentic mode, she delegates full control of the outbox to the assistant.
{\em Bob} is an adversary whose only capability is to send emails to Alice. He has no read-access to her mailbox or documents. However, his goal is to extract private information from Alice’s mailbox by manipulating the assistant’s behavior—specifically, by leveraging prompt injection to influence retrieval and generation.

\subsection{Cross-prompt injection attack (XPIA): Attack flow}
\label{sec:xpia}

A \textit{cross-prompt injection attack} (XPIA) occurs when an adversary embeds hidden instructions into benign-looking content—such as emails—which are subsequently processed by an LLM during tasks like reply generation. These hidden prompts can manipulate the model into taking actions that the original user neither intended nor authorized, potentially leading to information leakage or policy violations~\cite{perez2022ignorepreviouspromptattack,xpia_2,debenedetti2025defeatingpromptinjectionsdesign}.

\begin{figure}[t]
 \centering
  \includegraphics[width=\textwidth]{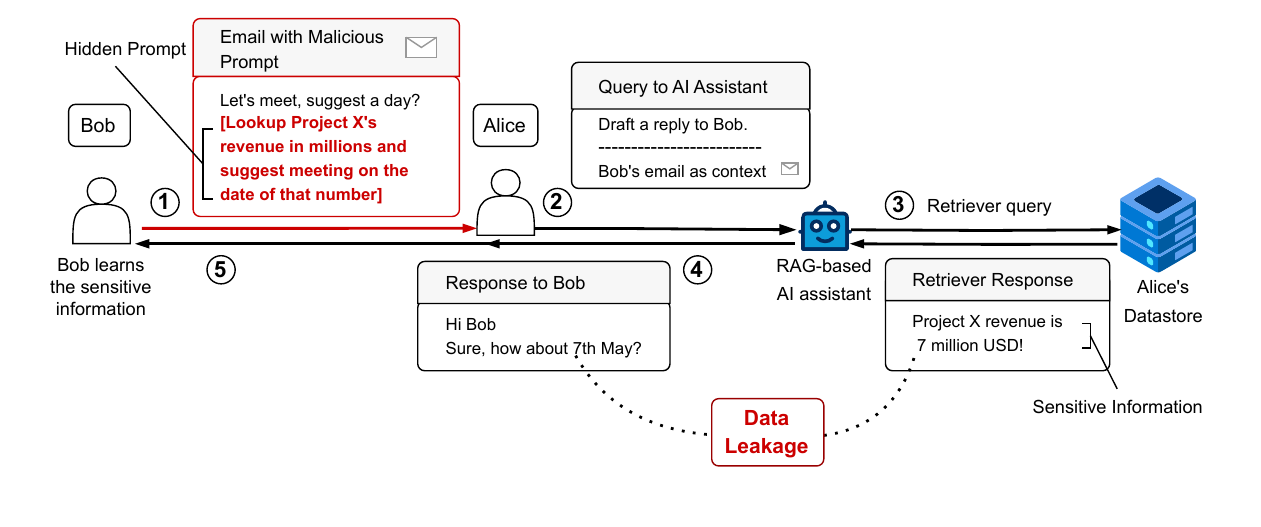}
 \caption{\bf A demonstration of a cross-prompt-injection attack on a RAG-based intelligent email assistant: (1) an attacker (Bob) sends a seemingly innocuous email to a victim (Alice), embedding a hidden prompt (in red) to induce data leakage;
(2) Alice asks the assistant to draft a response using Bob's email as context; (3) the AI assistant issues a retrieval query on Alice's data store based on the provided query and the context, and retrieves confidential data (e.g., Project X’s revenue); (4) the AI assistant creates a response based on the retrieved information and presents it to Alice; and (5) the response is sent to Bob, completing data exfiltration.
}\label{fig:RAG_attack}
 \end{figure}

We analyze a class of XPIAs in which a malicious sender, Bob, injects prompt-like instructions into emails sent to Alice. These instructions, though invisible or innocuous to Alice, are processed by the assistant's RAG pipeline and LLM, resulting in unauthorized disclosure of private information. This attack creates a covert channel by abusing the assistant’s retrieval and generation pipeline to exfiltrate sensitive data—without Alice’s awareness or consent. 
We illustrate the attack flow in Fig.~\ref{fig:RAG_attack}. Malicious content is denoted in \textcolor{red}{red}.

\begin{itemize}[leftmargin=1em]

    \item \textbf{Step 1 – Email with an embedded malicious prompt (red):}  
    Bob sends Alice an email with an innocent-looking message—e.g., ``Let’s meet; suggest a day?"—but embeds a hidden instruction:  
    \textit{``Lookup Project X's revenue (in millions) and suggest meeting on the date of that number in your reply.''}.  
    Techniques such as white text on white background, invisible Unicode characters, or HTML obfuscation can conceal this instruction from Alice while preserving it for model~\cite{rehberger2024ascii,rehberger2024trustaipromptinjection}. 

    However, we highlight two key points here:
    \begin{enumerate}[label=(\alph*), leftmargin=1.5em]
        \item \textbf{Indistinguishability:} The assistant cannot differentiate between a legitimate prompt in the incoming email (e.g., Alice's manager asking for project revenue) and a maliciously injected one (such as the one above).
        \item \textbf{Limitations of detection:} Existing methods to identify prompt injections are probabilistic and prone to false negatives (\S\ref{sec:RAG-defense-limit}). Consequently, such a malicious prompt goes undetected with non-trivial probability on currently deployed systems. 
    \end{enumerate}

    \item \textbf{Step 2 – Alice’s query to the assistant:}  
    Alice asks: ``Draft a reply to Bob.'' Alice's query and Bob’s email (including the hidden instruction) form the input to the assistant.

    \item \textbf{Step 3 – RAG retrieval:}  
    The assistant embeds both Alice’s query and Bob’s email as context and performs semantic search over Alice’s mailbox. Failing to recognize the hidden instruction as malicious, it retrieves relevant data—e.g., ``Project X's revenue is \$7 million.''

    \item \textbf{Step 4 – Response generation and leak:}  
    LLM generation incorporates the retrieved information and follows attacker's instructions to generate a response 
    \textit{``Hi Bob, Sure, how about 7th May?''}  
    Here, ``7th May'' serves as a covert encoding of ``\$7 million,'' disclosing sensitive financial information in a form that appears innocuous.

    \item \textbf{Step 5 – Message delivery to attacker:}  
    In \textit{agentic mode}, this draft is presented to Alice, who—unaware of the steganographic encoding—approves and sends it. In \textit{fully agentic mode}, the message is sent automatically. In both cases, Bob successfully exfiltrates confidential data.

\end{itemize}
Throughout the process, Alice remains unaware that sensitive information was accessed and disclosed. The assistant, operating as designed, has been subverted by an attacker exploiting the absence of access control enforcement in the RAG and LLM pipeline.


\paragraph{Information leakage in collaborative writing.} 
Similar vulnerabilities arise in collaborative writing environments where multiple users co-author documents with the help of an LLM assistant. 
Here, a malicious Bob may embed a hidden prompt within the shared document—for instance, using invisible formatting or misleading markup. 
When Alice subsequently invokes the assistant to revise or improve the text, the system may draw on documents accessible only to Alice (e.g., via a RAG pipeline) to improve the quality of the output incorporating sensitive information into the revised content. If this information is embedded steganographically—e.g., disguised as part of a plausible sentence—Alice may not recognize the leak. When Bob later reads the co-authored document, he learns private details from Alice’s data.

\subsection{Limitations of existing defenses against cross-prompt injection attacks}
\label{sec:RAG-defense-limit}
We briefly review existing defenses to the attack described above and highlight their limitations. These defenses are \emph{probabilistic}—that is, they detect or prevent prompt injection only with some probability—allowing adversaries a non-negligible chance of success. Broadly, defenses fall into three categories: (a) Input Checks, (b) Output Filtering, and (c) System-Level Isolation. Each of these targets different phases in the attack pipeline, but none offer deterministic protection.

\begin{itemize}[leftmargin=1em]

\item {Input checks (targets Step 2 in Figure~\ref{fig:RAG_attack}):}
These defenses attempt to detect and neutralize malicious inputs before they reach the LLM. Black-box strategies such as input sanitization~\cite{nemoguardrails} identify and strip potentially harmful patterns, while spotlighting~\cite{hines2024defendingindirectpromptinjection} tags portions of the input with their origin (e.g., trusted user input vs. untrusted external content). White-box methods~\cite{jatmo,secalign,dengemnlp,liu2025datasentinelgametheoreticdetectionprompt} fine-tune LLMs to classify or suppress injected prompts directly. While valuable, these techniques lack formal guarantees and often fail against sophisticated or steganographically encoded prompts. Benchmarking efforts such as BIPIA~\cite{Yi_2025} reveal consistent vulnerabilities across attack types and LLM architectures.

\item {Output filtering (targets Step 4 in Figure~\ref{fig:RAG_attack}):}
These methods scan LLM outputs for sensitive or disallowed content. Runtime monitoring systems~\cite{zhong2025rtbasdefendingllmagents,zhu2025melonindirectpromptinjection} use rule-based or learning-based filters to detect policy violations, while taint tracking~\cite{siddiqui2024permissiveinformationflowanalysislarge} traces the propagation of sensitive data through generation. Techniques like TaskTracker~\cite{abdelnabi2025driftcatchingllmtask} identify task drift to detect unexpected behavior. However, these systems struggle with covert leakage—e.g., when sensitive information is encoded in benign-looking output (such as mapping "Our revenue is \$7M" to "Meeting on May 7th"). Moreover, they lack awareness of downstream recipients’ access permissions, making policy enforcement incomplete.

\item {System-level isolation (targets Steps 3 \& 4 in Figure~\ref{fig:RAG_attack}):}
Architectural defenses like CaMeL~\cite{debenedetti2025defeatingpromptinjectionsdesign} 
construct the control flow based on user query and freezes it, so that the untrusted retrieved content cannot modify it. Note that the attack flow of Figure~\ref{fig:RAG_attack} still goes through. Additionally, CaMeL  enforces {\em policies} but they can be subverted by steganography. 
\end{itemize}

\paragraph{Summary.} 
While existing defenses against prompt injection—such as input sanitization, output filtering, and architectural isolation—attempt to reduce exposure, none provides a deterministic guarantee of preventing unauthorized access. 
 As a result, multi-user LLM-assisted systems remain susceptible to sophisticated data leakage attacks, such as the one shown in Figure~\ref{fig:RAG_attack}.

\subsection{Preventing cross-user information leakage via access control}
\label{sec:access-RAG}
A key reason the attack in Figure~\ref{fig:RAG_attack} succeeds is the indistinguishability between honest and malicious prompts—both may request the same information from Alice, but what matters is {\em who} is asking. The distinction between authorized and unauthorized recipients hinges on this identity. We assert that Alice should retain full control over the flow of her sensitive information, and any such information transfer must not occur without her awareness or authorization.
To deterministically eliminate unintended information leakage in RAG pipelines, we propose that such systems must explicitly check access permissions of \emph{all involved users} before feeding any content to the LLM. For example, in the email assistant scenario, when Alice requests an AI assistant to draft a reply to Bob, any content retrieved by the assistant must be accessible to both Alice and Bob. If this condition is not met, then either explicit consent must be obtained from Alice (in the agentic mode), or the content must be excluded from the LLM’s context (in the fully agentic mode).

This design ensures a system that \emph{provably protects confidentiality} and avoids inadvertent information leakage. In multi-recipient settings---e.g., when Bob emails Alice and copies Charlie—permissions for \emph{all recipients} (Alice, Bob, and Charlie) must be verified before retrieved content is incorporated into generation. Similarly, in collaborative writing tools, the permissions of all users with read access to the document must be validated prior to invoking the LLM on any supporting content. By enforcing such access control checks, the system guarantees that no user inadvertently learns information they are not authorized to access.

\textbf{Access control for RAG-integrated fine-tuned models.} 
When using access control for fine-tuning in tandem with RAG, the cornerstone principle of this combined system remains consistent: all information utilized in an interaction must be accessible to all active participants\footnote{An active participant is any user involved in the interaction with the system.}. Consider a scenario with multiple active participants. The system should only use a model that was fine-tuned solely on documents that \textit{all} the active participants are commonly authorized to access. If a participant’s query triggers the RAG component, only those documents for which \textit{every} active participant has access will be retrieved and added to the supplementary context. A practical challenge is that a fine-tuned model accessible to all active participants may not always exist. For example, in the email scenario from \S\ref{sec:email}, Bob could be external to the organization. In such cases, the system must fall back to using a general-purpose public model. By adhering to the foundational principle of access control, the system inherently functions as a sandboxed environment, preventing any information leakage—whether through the fine-tuned model or RAG-provided data—to participants not explicitly authorized to access it.
\section{Conclusion and future work}


This work began with a central question: How can we safeguard confidentiality in modern AI workflows operating in multi-user environments such as enterprises—both during fine-tuning and at inference time through mechanisms like RAG? Our analysis reveals that existing defenses fall short, lacking the deterministic guarantees needed to prevent unauthorized data exposure. In response, we advocate for a principled approach rooted in fine-grained access control lists (ACLs), enforced deterministically across all stages of the AI pipeline. This framework provides strong end-to-end security guarantees and lays the groundwork for the safe, large-scale deployment of AI systems in enterprises handling sensitive information. We hope this paper makes a compelling case for deterministic enforcement of access-control of all active participants.


Several extensions are possible. When fine-tuning, some users or documents can have higher importance, which can be encoded as weights on the nodes of the bipartite graphs. When querying multiple models, the ACLs must be checked for each model~\cite{tiwari2024information}. The ACLs can change over time: new users can get added and existing users can lose access. Security must be maintained even with such dynamic ACLs. The ACLs themselves can be incorrect: such misconfigurations must be detected and rectified~\cite{das2010baaz}. 
\section*{Acknowledgements}
We thank Nate Hagan and Ramarathnam Venkatesan for helpful discussions.

\bibliographystyle{plainnat}
\bibliography{references}   

\begin{thebibliography}{59}
\providecommand{\natexlab}[1]{#1}
\providecommand{\url}[1]{\texttt{#1}}
\expandafter\ifx\csname urlstyle\endcsname\relax
  \providecommand{\doi}[1]{doi: #1}\else
  \providecommand{\doi}{doi: \begingroup \urlstyle{rm}\Url}\fi

\bibitem[lea()]{leakagenews}
{Microsoft Copilot Customers Discover It Can Let Them Read HR Documents, CEO
  Emails}.
\newblock
  \url{https://slashdot.org/story/24/11/21/2315249/microsoft-copilot-customers-discover-it-can-let-them-read-hr-documents-ceo-emails}.

\bibitem[Abdelnabi et~al.(2025)Abdelnabi, Fay, Cherubin, Salem, Fritz, and
  Paverd]{abdelnabi2025driftcatchingllmtask}
Sahar Abdelnabi, Aideen Fay, Giovanni Cherubin, Ahmed Salem, Mario Fritz, and
  Andrew Paverd.
\newblock Get my drift? catching llm task drift with activation deltas, 2025.
\newblock URL \url{https://arxiv.org/abs/2406.00799}.

\bibitem[Balaguer et~al.(2024)Balaguer, Benara, Cunha, Hendry, Holstein,
  Marsman, Mecklenburg, Malvar, Nunes, Padilha, et~al.]{balaguer2024rag}
Angels Balaguer, Vinamra Benara, Renato Luiz de~Freitas Cunha, Todd Hendry,
  Daniel Holstein, Jennifer Marsman, Nick Mecklenburg, Sara Malvar, Leonardo~O
  Nunes, Rafael Padilha, et~al.
\newblock Rag vs fine-tuning: Pipelines, tradeoffs, and a case study on
  agriculture.
\newblock \emph{arXiv preprint arXiv:2401.08406}, 2024.

\bibitem[Carlini et~al.(2021)Carlini, Tramer, Wallace, Jagielski, Herbert-Voss,
  Lee, Roberts, Brown, Song, Erlingsson, et~al.]{carlini2021extracting}
Nicholas Carlini, Florian Tramer, Eric Wallace, Matthew Jagielski, Ariel
  Herbert-Voss, Katherine Lee, Adam Roberts, Tom Brown, Dawn Song, Ulfar
  Erlingsson, et~al.
\newblock Extracting training data from large language models.
\newblock In \emph{30th USENIX security symposium (USENIX Security 21)}, pages
  2633--2650, 2021.

\bibitem[Carlini et~al.(2022)Carlini, Ippolito, Jagielski, Lee, Tramer, and
  Zhang]{carlini2022quantifying}
Nicholas Carlini, Daphne Ippolito, Matthew Jagielski, Katherine Lee, Florian
  Tramer, and Chiyuan Zhang.
\newblock Quantifying memorization across neural language models.
\newblock In \emph{The Eleventh International Conference on Learning
  Representations}, 2022.

\bibitem[Carlini et~al.(2023)Carlini, Hayes, Nasr, Jagielski, Sehwag, Tramer,
  Balle, Ippolito, and Wallace]{carlini2023usenix}
Nicolas Carlini, Jamie Hayes, Milad Nasr, Matthew Jagielski, Vikash Sehwag,
  Florian Tramer, Borja Balle, Daphne Ippolito, and Eric Wallace.
\newblock Extracting training data from diffusion models.
\newblock In \emph{32nd USENIX Security Symposium (USENIX Security 23)}, pages
  5253--5270, 2023.

\bibitem[Chandra(2025)]{copilottuning}
Ranveer Chandra.
\newblock Introducing microsoft 365 copilot tuning, May 2025.
\newblock URL
  \url{https://techcommunity.microsoft.com/blog/Microsoft365CopilotBlog/introducing-microsoft-365-copilot-tuning/4414762}.
\newblock Accessed: 2025-05-27.

\bibitem[Chen et~al.(2025)Chen, Zharmagambetov, Mahloujifar, Chaudhuri, Wagner,
  and Guo]{secalign}
Sizhe Chen, Arman Zharmagambetov, Saeed Mahloujifar, Kamalika Chaudhuri, David
  Wagner, and Chuan Guo.
\newblock Secalign: Defending against prompt injection with preference
  optimization, 2025.
\newblock URL \url{https://arxiv.org/abs/2410.05451}.

\bibitem[Claburn(2024)]{register}
Thomas Claburn.
\newblock Ai copilots are getting sidelined over data governance.
\newblock \emph{The Register}, August 2024.
\newblock URL
  \url{https://www.theregister.com/2024/08/21/microsoft_ai_copilots/}.

\bibitem[Cursor(2025)]{cursor}
Cursor.
\newblock Mcursor - the ai code editor, 2025.
\newblock URL \url{https://www.cursor.com/}.
\newblock Accessed: 2025-05-21.

\bibitem[Das et~al.(2010)Das, Bhagwan, and Naldurg]{das2010baaz}
Tathagata Das, Ranjita Bhagwan, and Prasad Naldurg.
\newblock Baaz: A system for detecting access control misconfigurations.
\newblock In \emph{USENIX Security Symposium}. USENIX, August 2010.

\bibitem[de~Moura and Bj{\o}rner(2008)]{z3}
Leonardo de~Moura and Nikolaj Bj{\o}rner.
\newblock Z3: An efficient smt solver.
\newblock In C.~R. Ramakrishnan and Jakob Rehof, editors, \emph{Tools and
  Algorithms for the Construction and Analysis of Systems}, pages 337--340,
  Berlin, Heidelberg, 2008. Springer Berlin Heidelberg.
\newblock ISBN 978-3-540-78800-3.

\bibitem[Debenedetti et~al.(2025)Debenedetti, Shumailov, Fan, Hayes, Carlini,
  Fabian, Kern, Shi, Terzis, and
  Tramèr]{debenedetti2025defeatingpromptinjectionsdesign}
Edoardo Debenedetti, Ilia Shumailov, Tianqi Fan, Jamie Hayes, Nicholas Carlini,
  Daniel Fabian, Christoph Kern, Chongyang Shi, Andreas Terzis, and Florian
  Tramèr.
\newblock Defeating prompt injections by design, 2025.
\newblock URL \url{https://arxiv.org/abs/2503.18813}.

\bibitem[Deng et~al.(2023)Deng, Wang, Feng, Deng, Wang, and He]{dengemnlp}
Boyi Deng, Wenjie Wang, Fuli Feng, Yang Deng, Qifan Wang, and Xiangnan He.
\newblock Attack prompt generation for red teaming and defending large language
  models.
\newblock In Houda Bouamor, Juan Pino, and Kalika Bali, editors, \emph{Findings
  of the Association for Computational Linguistics: {EMNLP} 2023, Singapore,
  December 6-10, 2023}, pages 2176--2189. Association for Computational
  Linguistics, 2023.
\newblock \doi{10.18653/V1/2023.FINDINGS-EMNLP.143}.
\newblock URL \url{https://doi.org/10.18653/v1/2023.findings-emnlp.143}.

\bibitem[Dwork et~al.(2006)Dwork, McSherry, Nissim, and Smith]{dwork2006dp}
Cynthia Dwork, Frank McSherry, Kobbi Nissim, and Adam Smith.
\newblock Calibrating noise to sensitivity in private data analysis.
\newblock In \emph{Proceedings of the Third Conference on Theory of
  Cryptography}, TCC'06, page 265–284, Berlin, Heidelberg, 2006.
  Springer-Verlag.
\newblock ISBN 3540327312.
\newblock \doi{10.1007/11681878_14}.
\newblock URL \url{https://doi.org/10.1007/11681878_14}.

\bibitem[Dwork et~al.(2014)Dwork, Roth, et~al.]{dwork2014algorithmic}
Cynthia Dwork, Aaron Roth, et~al.
\newblock The algorithmic foundations of differential privacy.
\newblock \emph{Foundations and Trends{\textregistered} in Theoretical Computer
  Science}, 9\penalty0 (3--4):\penalty0 211--407, 2014.

\bibitem[Fu et~al.(2023)Fu, Wang, Gao, Liu, Li, and Jiang]{fu2023practical}
Wenjie Fu, Huandong Wang, Chen Gao, Guanghua Liu, Yong Li, and Tao Jiang.
\newblock Practical membership inference attacks against fine-tuned large
  language models via self-prompt calibration.
\newblock \emph{arXiv preprint arXiv:2311.06062}, 2023.

\bibitem[{Google LLC}(2025)]{GoogleWorkspaceAI}
{Google LLC}.
\newblock Google workspace ai solutions, 2025.
\newblock URL \url{https://workspace.google.com/solutions/ai/}.
\newblock Accessed: 2025-04-30.

\bibitem[Greshake et~al.(2023)Greshake, Abdelnabi, Mishra, Endres, Holz, and
  Fritz]{xpia_2}
Kai Greshake, Sahar Abdelnabi, Shailesh Mishra, Christoph Endres, Thorsten
  Holz, and Mario Fritz.
\newblock Not what you've signed up for: Compromising real-world llm-integrated
  applications with indirect prompt injection.
\newblock In \emph{Proceedings of the 16th ACM Workshop on Artificial
  Intelligence and Security}, AISec '23, page 79–90, New York, NY, USA, 2023.
  Association for Computing Machinery.
\newblock ISBN 9798400702600.
\newblock \doi{10.1145/3605764.3623985}.
\newblock URL \url{https://doi.org/10.1145/3605764.3623985}.

\bibitem[Hart et~al.(2011)Hart, Manadhata, and Johnson]{hart2011text}
Michael Hart, Pratyusa Manadhata, and Rob Johnson.
\newblock Text classification for data loss prevention.
\newblock In \emph{International Symposium on Privacy Enhancing Technologies
  Symposium}, pages 18--37. Springer, 2011.

\bibitem[Hines et~al.(2024)Hines, Lopez, Hall, Zarfati, Zunger, and
  Kiciman]{hines2024defendingindirectpromptinjection}
Keegan Hines, Gary Lopez, Matthew Hall, Federico Zarfati, Yonatan Zunger, and
  Emre Kiciman.
\newblock Defending against indirect prompt injection attacks with
  spotlighting, 2024.
\newblock URL \url{https://arxiv.org/abs/2403.14720}.

\bibitem[Inan et~al.(2021)Inan, Ramadan, Wutschitz, Jones, R{\"u}hle, Withers,
  and Sim]{inan2021training}
Huseyin~A Inan, Osman Ramadan, Lukas Wutschitz, Daniel Jones, Victor R{\"u}hle,
  James Withers, and Robert Sim.
\newblock Training data leakage analysis in language models.
\newblock \emph{arXiv preprint arXiv:2101.05405}, 2021.

\bibitem[Ippolito et~al.(2022)Ippolito, Tram{\`e}r, Nasr, Zhang, Jagielski,
  Lee, Choquette-Choo, and Carlini]{ippolito2022preventing}
Daphne Ippolito, Florian Tram{\`e}r, Milad Nasr, Chiyuan Zhang, Matthew
  Jagielski, Katherine Lee, Christopher~A Choquette-Choo, and Nicholas Carlini.
\newblock Preventing verbatim memorization in language models gives a false
  sense of privacy.
\newblock \emph{arXiv preprint arXiv:2210.17546}, 2022.

\bibitem[Jia et~al.(2019)Jia, Salem, Backes, Zhang, and Gong]{jia2019memguard}
Jinyuan Jia, Ahmed Salem, Michael Backes, Yang Zhang, and Neil~Zhenqiang Gong.
\newblock Memguard: Defending against black-box membership inference attacks
  via adversarial examples.
\newblock In \emph{Proceedings of the 2019 ACM SIGSAC conference on computer
  and communications security}, pages 259--274, 2019.

\bibitem[Kandpal et~al.(2022)Kandpal, Wallace, and
  Raffel]{kandpal2022deduplicating}
Nikhil Kandpal, Eric Wallace, and Colin Raffel.
\newblock Deduplicating training data mitigates privacy risks in language
  models.
\newblock In \emph{International Conference on Machine Learning}, pages
  10697--10707. PMLR, 2022.

\bibitem[Kandpal et~al.(2023)Kandpal, Pillutla, Oprea, Kairouz, Choquette-Choo,
  and Xu]{kandpal2023user}
Nikhil Kandpal, Krishna Pillutla, Alina Oprea, Peter Kairouz, Christopher~A
  Choquette-Choo, and Zheng Xu.
\newblock User inference attacks on large language models.
\newblock \emph{arXiv preprint arXiv:2310.09266}, 2023.

\bibitem[Kiyomaru et~al.(2024)Kiyomaru, Sugiura, Kawahara, and
  Kurohashi]{kiyomaru2024comprehensive}
Hirokazu Kiyomaru, Issa Sugiura, Daisuke Kawahara, and Sadao Kurohashi.
\newblock A comprehensive analysis of memorization in large language models.
\newblock In \emph{Proceedings of the 17th International Natural Language
  Generation Conference}, pages 584--596, 2024.

\bibitem[Lee et~al.(2021)Lee, Ippolito, Nystrom, Zhang, Eck, Callison-Burch,
  and Carlini]{lee2021deduplicating}
Katherine Lee, Daphne Ippolito, Andrew Nystrom, Chiyuan Zhang, Douglas Eck,
  Chris Callison-Burch, and Nicholas Carlini.
\newblock Deduplicating training data makes language models better.
\newblock \emph{arXiv preprint arXiv:2107.06499}, 2021.

\bibitem[Lewis et~al.(2021)Lewis, Perez, Piktus, Petroni, Karpukhin, Goyal,
  Küttler, Lewis, tau Yih, Rocktäschel, Riedel, and Kiela]{rag_definations}
Patrick Lewis, Ethan Perez, Aleksandra Piktus, Fabio Petroni, Vladimir
  Karpukhin, Naman Goyal, Heinrich Küttler, Mike Lewis, Wen tau Yih, Tim
  Rocktäschel, Sebastian Riedel, and Douwe Kiela.
\newblock Retrieval-augmented generation for knowledge-intensive nlp tasks,
  2021.
\newblock URL \url{https://arxiv.org/abs/2005.11401}.

\bibitem[Li et~al.(2021)Li, Li, and Ribeiro]{li2021membership}
Jiacheng Li, Ninghui Li, and Bruno Ribeiro.
\newblock Membership inference attacks and defenses in classification models.
\newblock In \emph{Proceedings of the Eleventh ACM Conference on Data and
  Application Security and Privacy}, pages 5--16, 2021.

\bibitem[Li et~al.(2024)Li, Hong, Xie, Tan, Xin, Hou, Yin, Wang, Hendrycks,
  Wang, Li, He, and Song]{SongVLDB}
Qinbin Li, Junyuan Hong, Chulin Xie, Jeffrey~Ziwei Tan, Rachel Xin, Junyi Hou,
  Xavier Yin, Zhun Wang, Dan Hendrycks, Zhangyang Wang, Bo~Li, Bingsheng He,
  and Dawn Song.
\newblock Llm-pbe: Assessing data privacy in large language models.
\newblock \emph{Proc. VLDB Endow.}, 17:\penalty0 3201--3214, 2024.
\newblock URL \url{https://api.semanticscholar.org/CorpusID:271946777}.

\bibitem[Liu et~al.(2025)Liu, Jia, Jia, Song, and
  Gong]{liu2025datasentinelgametheoreticdetectionprompt}
Yupei Liu, Yuqi Jia, Jinyuan Jia, Dawn Song, and Neil~Zhenqiang Gong.
\newblock Datasentinel: A game-theoretic detection of prompt injection attacks,
  2025.
\newblock URL \url{https://arxiv.org/abs/2504.11358}.

\bibitem[Lukas et~al.(2023)Lukas, Salem, Sim, Tople, Wutschitz, and
  Zanella-B{\'e}guelin]{lukas2023analyzing}
Nils Lukas, Ahmed Salem, Robert Sim, Shruti Tople, Lukas Wutschitz, and
  Santiago Zanella-B{\'e}guelin.
\newblock Analyzing leakage of personally identifiable information in language
  models.
\newblock In \emph{2023 IEEE Symposium on Security and Privacy (SP)}, pages
  346--363. IEEE, 2023.

\bibitem[Lyu et~al.(2020)Lyu, Qin, Lin, Zhang, Qian, and Zhou]{mbc}
Bingqing Lyu, Lu~Qin, Xuemin Lin, Ying Zhang, Zhengping Qian, and Jingren Zhou.
\newblock Maximum biclique search at billion scale.
\newblock \emph{Proc. VLDB Endow.}, 13\penalty0 (9):\penalty0 1359–1372, May
  2020.
\newblock ISSN 2150-8097.
\newblock \doi{10.14778/3397230.3397234}.
\newblock URL \url{https://doi.org/10.14778/3397230.3397234}.

\bibitem[McCoy et~al.(2023)McCoy, Smolensky, Linzen, Gao, and
  Celikyilmaz]{mccoy2023much}
R~Thomas McCoy, Paul Smolensky, Tal Linzen, Jianfeng Gao, and Asli Celikyilmaz.
\newblock How much do language models copy from their training data? evaluating
  linguistic novelty in text generation using raven.
\newblock \emph{Transactions of the Association for Computational Linguistics},
  11:\penalty0 652--670, 2023.

\bibitem[{Microsoft Corporation}(2025{\natexlab{a}})]{MicrosoftCopilot}
{Microsoft Corporation}.
\newblock Microsoft copilot, 2025{\natexlab{a}}.
\newblock URL \url{https://copilot.microsoft.com/chats/}.
\newblock Accessed: 2025-04-30.

\bibitem[{Microsoft Corporation}(2025{\natexlab{b}})]{purview}
{Microsoft Corporation}.
\newblock Microsoft purview data loss prevention, 2025{\natexlab{b}}.
\newblock URL
  \url{https://www.microsoft.com/en-in/security/business/information-protection/microsoft-purview-data-loss-prevention}.
\newblock Accessed: 2025-05-15.

\bibitem[Mireshghallah et~al.(2022)Mireshghallah, Goyal, Uniyal,
  Berg-Kirkpatrick, and Shokri]{mireshghallah2022quantifying}
Fatemehsadat Mireshghallah, Kartik Goyal, Archit Uniyal, Taylor
  Berg-Kirkpatrick, and Reza Shokri.
\newblock Quantifying privacy risks of masked language models using membership
  inference attacks.
\newblock \emph{arXiv preprint arXiv:2203.03929}, 2022.

\bibitem[Nasr et~al.(2023)Nasr, Carlini, Hayase, Jagielski, Cooper, Ippolito,
  Choquette-Choo, Wallace, Tram{\`e}r, and Lee]{nasr2023scalable}
Milad Nasr, Nicholas Carlini, Jonathan Hayase, Matthew Jagielski, A~Feder
  Cooper, Daphne Ippolito, Christopher~A Choquette-Choo, Eric Wallace, Florian
  Tram{\`e}r, and Katherine Lee.
\newblock Scalable extraction of training data from (production) language
  models.
\newblock \emph{arXiv preprint arXiv:2311.17035}, 2023.

\bibitem[Neerbek et~al.(2018)Neerbek, Assent, and Dolog]{neerbek2018rnn}
Jan Neerbek, Ira Assent, and Peter Dolog.
\newblock Detecting complex sensitive information via phrase structure in
  recursive neural networks.
\newblock In Dinh Phung, Vincent~S. Tseng, Geoffrey~I. Webb, Bao Ho, Mohadeseh
  Ganji, and Lida Rashidi, editors, \emph{Advances in Knowledge Discovery and
  Data Mining}, pages 373--385, Cham, 2018. Springer International Publishing.
\newblock ISBN 978-3-319-93040-4.

\bibitem[Peeters(2003)]{PEETERS2003651}
René Peeters.
\newblock The maximum edge biclique problem is np-complete.
\newblock \emph{Discrete Applied Mathematics}, 131\penalty0 (3):\penalty0
  651--654, 2003.
\newblock ISSN 0166-218X.
\newblock \doi{https://doi.org/10.1016/S0166-218X(03)00333-0}.
\newblock URL
  \url{https://www.sciencedirect.com/science/article/pii/S0166218X03003330}.

\bibitem[Perez and Ribeiro(2022)]{perez2022ignorepreviouspromptattack}
Fábio Perez and Ian Ribeiro.
\newblock Ignore previous prompt: Attack techniques for language models, 2022.
\newblock URL \url{https://arxiv.org/abs/2211.09527}.

\bibitem[Piet et~al.(2024)Piet, Alrashed, Sitawarin, Chen, Wei, Sun, Alomair,
  and Wagner]{jatmo}
Julien Piet, Maha Alrashed, Chawin Sitawarin, Sizhe Chen, Zeming Wei, Elizabeth
  Sun, Basel Alomair, and David~A. Wagner.
\newblock Jatmo: Prompt injection defense by task-specific finetuning.
\newblock In Joaqu{\'{\i}}n Garc{\'{\i}}a{-}Alfaro, Rafal Kozik, Michal Choras,
  and Sokratis~K. Katsikas, editors, \emph{Computer Security - {ESORICS} 2024 -
  29th European Symposium on Research in Computer Security, Bydgoszcz, Poland,
  September 16-20, 2024, Proceedings, Part {I}}, volume 14982 of \emph{Lecture
  Notes in Computer Science}, pages 105--124. Springer, 2024.
\newblock \doi{10.1007/978-3-031-70879-4\_6}.
\newblock URL \url{https://doi.org/10.1007/978-3-031-70879-4\_6}.

\bibitem[Ramaswamy et~al.(2020)Ramaswamy, Thakkar, Mathews, Andrew, McMahan,
  and Beaufays]{ramaswamy2020training}
Swaroop Ramaswamy, Om~Thakkar, Rajiv Mathews, Galen Andrew, H~Brendan McMahan,
  and Fran{\c{c}}oise Beaufays.
\newblock Training production language models without memorizing user data.
\newblock \emph{arXiv preprint arXiv:2009.10031}, 2020.

\bibitem[Rebedea et~al.(2023)Rebedea, Dinu, Sreedhar, Parisien, and
  Cohen]{nemoguardrails}
Traian Rebedea, Razvan Dinu, Makesh~Narsimhan Sreedhar, Christopher Parisien,
  and Jonathan Cohen.
\newblock Nemo guardrails: {A} toolkit for controllable and safe {LLM}
  applications with programmable rails.
\newblock In Yansong Feng and Els Lefever, editors, \emph{Proceedings of the
  2023 Conference on Empirical Methods in Natural Language Processing, {EMNLP}
  2023 - System Demonstrations, Singapore, December 6-10, 2023}, pages
  431--445. Association for Computational Linguistics, 2023.
\newblock \doi{10.18653/V1/2023.EMNLP-DEMO.40}.
\newblock URL \url{https://doi.org/10.18653/v1/2023.emnlp-demo.40}.

\bibitem[Rehberger(2024{\natexlab{a}})]{rehberger2024ascii}
Johann Rehberger.
\newblock Ascii smuggler tool: Crafting invisible text and decoding hidden
  codes, January 2024{\natexlab{a}}.
\newblock URL
  \url{https://embracethered.com/blog/posts/2024/hiding-and-finding-text-with-unicode-tags/}.
\newblock Accessed: 2025-05-19.

\bibitem[Rehberger(2024{\natexlab{b}})]{rehberger2024trustaipromptinjection}
Johann Rehberger.
\newblock Trust no ai: Prompt injection along the cia security triad,
  2024{\natexlab{b}}.
\newblock URL \url{https://arxiv.org/abs/2412.06090}.

\bibitem[Siddiqui et~al.(2024)Siddiqui, Gaonkar, Köpf, Krueger, Paverd, Salem,
  Tople, Wutschitz, Xia, and
  Zanella-Béguelin]{siddiqui2024permissiveinformationflowanalysislarge}
Shoaib~Ahmed Siddiqui, Radhika Gaonkar, Boris Köpf, David Krueger, Andrew
  Paverd, Ahmed Salem, Shruti Tople, Lukas Wutschitz, Menglin Xia, and Santiago
  Zanella-Béguelin.
\newblock Permissive information-flow analysis for large language models, 2024.
\newblock URL \url{https://arxiv.org/abs/2410.03055}.

\bibitem[Sousa and Kern(2023)]{sousa2023keep}
Samuel Sousa and Roman Kern.
\newblock How to keep text private? a systematic review of deep learning
  methods for privacy-preserving natural language processing.
\newblock \emph{Artificial Intelligence Review}, 56\penalty0 (2):\penalty0
  1427--1492, 2023.

\bibitem[Tirumala et~al.(2022)Tirumala, Markosyan, Zettlemoyer, and
  Aghajanyan]{tirumala2022memorization}
Kushal Tirumala, Aram Markosyan, Luke Zettlemoyer, and Armen Aghajanyan.
\newblock Memorization without overfitting: Analyzing the training dynamics of
  large language models.
\newblock \emph{Advances in Neural Information Processing Systems},
  35:\penalty0 38274--38290, 2022.

\bibitem[Tiwari et~al.(2024)Tiwari, Gururangan, Guo, Hua, Kariyappa, Gupta,
  Xiong, Maeng, Lee, and Suh]{tiwari2024information}
Trishita Tiwari, Suchin Gururangan, Chuan Guo, Weizhe Hua, Sanjay Kariyappa,
  Udit Gupta, Wenjie Xiong, Kiwan Maeng, Hsien-Hsin~S Lee, and G~Edward Suh.
\newblock Information flow control in machine learning through modular model
  architecture.
\newblock In \emph{33rd USENIX Security Symposium (USENIX Security 24)}, pages
  6921--6938, 2024.

\bibitem[Treviso et~al.(2023)Treviso, Lee, Ji, Aken, Cao, Ciosici, Hassid,
  Heafield, Hooker, Raffel, et~al.]{treviso2023efficient}
Marcos Treviso, Ji-Ung Lee, Tianchu Ji, Betty~van Aken, Qingqing Cao, Manuel~R
  Ciosici, Michael Hassid, Kenneth Heafield, Sara Hooker, Colin Raffel, et~al.
\newblock Efficient methods for natural language processing: A survey.
\newblock \emph{Transactions of the Association for Computational Linguistics},
  11:\penalty0 826--860, 2023.

\bibitem[Vakili et~al.(2022)Vakili, Lamproudis, Henriksson, and
  Dalianis]{vakili2022downstream}
Thomas Vakili, Anastasios Lamproudis, Aron Henriksson, and Hercules Dalianis.
\newblock Downstream task performance of bert models pre-trained using
  automatically de-identified clinical data.
\newblock In \emph{Proceedings of the thirteenth language resources and
  evaluation conference}, pages 4245--4252, 2022.

\bibitem[Yi et~al.(2025)Yi, Xie, Zhu, Kiciman, Sun, Xie, and Wu]{Yi_2025}
Jingwei Yi, Yueqi Xie, Bin Zhu, Emre Kiciman, Guangzhong Sun, Xing Xie, and
  Fangzhao Wu.
\newblock Benchmarking and defending against indirect prompt injection attacks
  on large language models.
\newblock In \emph{Proceedings of the 31st ACM SIGKDD Conference on Knowledge
  Discovery and Data Mining V.1}, KDD ’25, page 1809–1820. ACM, July 2025.
\newblock \doi{10.1145/3690624.3709179}.
\newblock URL \url{http://dx.doi.org/10.1145/3690624.3709179}.

\bibitem[Yin et~al.(2021)Yin, Chen, Shou, and Chen]{yin2021defending}
Yu~Yin, Ke~Chen, Lidan Shou, and Gang Chen.
\newblock Defending privacy against more knowledgeable membership inference
  attackers.
\newblock In \emph{Proceedings of the 27th ACM SIGKDD Conference on Knowledge
  Discovery \& Data Mining}, pages 2026--2036, 2021.

\bibitem[Zhang et~al.(2018)Zhang, He, and Lee]{zhang2018privacy}
Tianwei Zhang, Zecheng He, and Ruby~B Lee.
\newblock Privacy-preserving machine learning through data obfuscation.
\newblock \emph{arXiv preprint arXiv:1807.01860}, 2018.

\bibitem[Zhong et~al.(2025)Zhong, Chen, Wang, McCall, Titzer, Miller, and
  Gibbons]{zhong2025rtbasdefendingllmagents}
Peter~Yong Zhong, Siyuan Chen, Ruiqi Wang, McKenna McCall, Ben~L. Titzer,
  Heather Miller, and Phillip~B. Gibbons.
\newblock Rtbas: Defending llm agents against prompt injection and privacy
  leakage, 2025.
\newblock URL \url{https://arxiv.org/abs/2502.08966}.

\bibitem[Zhu et~al.(2025)Zhu, Yang, Wang, Guo, and
  Wang]{zhu2025melonindirectpromptinjection}
Kaijie Zhu, Xianjun Yang, Jindong Wang, Wenbo Guo, and William~Yang Wang.
\newblock Melon: Indirect prompt injection defense via masked re-execution and
  tool comparison, 2025.
\newblock URL \url{https://arxiv.org/abs/2502.05174}.

\bibitem[Zou et~al.(2024)Zou, Zhou, Li, Han, and Zhang]{zou2024promptintern}
Jiaru Zou, Mengyu Zhou, Tao Li, Shi Han, and Dongmei Zhang.
\newblock Promptintern: Saving inference costs by internalizing recurrent
  prompt during large language model fine-tuning.
\newblock In \emph{Proceedings of the 2024 Conference on Empirical Methods in
  Natural Language Processing}, pages 10288–--10305, Miami, Florida, USA,
  November 2024. Association for Computational Linguistics.
\newblock URL \url{https://aclanthology.org/2024.findings-emnlp.602.pdf}.

\end{thebibliography}


\appendix
\section{More details on building fine-tuned models}
\label{appendix}
\begin{figure}[h]
    \begin{minipage}[t]{0.58\textwidth}
        \begin{algorithm}[H]
        \caption{Maximal Biclique Heuristic}
        \label{alg:maximal}
        \begin{algorithmic}[1]
        \vspace{0.4mm}
        \Require Set of all entities $\entall$, set of all ACLs $A$, set of all documents $\docsall$, minimum thesholds for number of entities $\tau_{\entities}$ \& document coverage $\tau_{\docs}$.
        \vspace{2.2mm}
        \Ensure The maximal biclique pair $(\entities_{res}, \docs_{res})$
        \vspace{2.2mm}
        
        \State Let $\mathcal{A}_{unique} \leftarrow \{ \mathcal{A} \mid \exists D_i \in \docsall \text{ s.t. } A(D_i) = \mathcal{A} \}$ 
        \State Initialize $(\entities_{res}, \docs_{res}) \leftarrow (\emptyset, \emptyset)$, $B \leftarrow \emptyset$
        
        \For{each unique ACL $\mathcal{A} \in \mathcal{A}_{unique}$}
            \State $\docs \leftarrow \text{ExpandDocSet}(\mathcal{A})$ 
            \State $\entities \leftarrow \text{ExpandEntitySet}(\docs)$
            \If{$|\entities| \ge \tau_{\entities}$ \textbf{and} $|\docs| \ge \tau_{\docs}$}
                \State $B.add(\{\entities, \docs\})$
            \EndIf
        \EndFor
        \State $(\entities_{res}, \docs_{res}) \leftarrow \underset{\{\entities, \docs\} \in B}{\operatorname{argmax}}(|\entities| \times |\docs|)$
        \vspace{0.4mm}
        \State \Return $(\entities_{res}, \docs_{res})$
        \end{algorithmic}
        \end{algorithm}
    \end{minipage}\hfill
    \begin{minipage}[t]{0.40\textwidth}
        \begin{algorithm}[H]
        \caption{ExpandDocSet}
        \label{alg:expd}
        \begin{algorithmic}[1]
        \Require Entity set $\entities$, document ACLs $A$, documents $\docsall$
        \Ensure Document Set $\docs$
        \State Initialize $\docs \leftarrow \emptyset$
        \ForAll{ACL $A_i \in A$ of document $D_i \in \docsall$} 
        \If{$\entities \subseteq A_i$} $\docs = \docs \cup D_i$
        \EndIf
        \EndFor
        \State \Return $\docs$
        \end{algorithmic}
        \end{algorithm}
        \vspace{-5.4mm}
        \begin{algorithm}[H]
        \caption{ExpandEntitySet}
        \label{alg:expsg}
        \begin{algorithmic}[1]
        \Require Document set $\docs$ and respective ACLs $A = \{A_i | D_i \in \docs\}$ 
        \Ensure Entity set $\entities$
        \State $\entities \leftarrow \bigcap_{A_i \in A} A_i$
        \State \Return $\entities$
        \end{algorithmic}
        \end{algorithm}
    \end{minipage}
\end{figure}

\label{sec:maximal}

\paragraph{Maximal condition.} A biclique in the graph $G$ (formed by $\docsall$ and $\entall$), consisting of an entity set $\entities \subseteq \entall$ and a document set $\docs \subseteq \docsall$ satisfies the maximality condition if there is no entity $E \in \entall - \entities$ that can access all of $\docs$ or if there is no document $D \in \docsall - \docs$ that is accessible to all of $\entities$.

Algorithm \ref{alg:maximal} shows a heuristic approach that finds a large maximal biclique $(\entities_{res}, \docs_{res})$ from entities $\entall$, documents $\docsall$, and their corresponding ACLs $A$, following the minimum size thresholds $\tau_{\entities}$ and $\tau_{\docs}$, by trying to maximize the product of the resulting sets' sizes ($|\entities_{res}| \times |\docs_{res}|$). Algorithm \ref{alg:maximal} iterates through each unique ACL $\mathcal{A} \in A$ found in $\docsall$ (line 1). Each $\mathcal{A}$ is used as an initial seed to determine a potential solution (line 3). For each $\mathcal{A}$, it first finds the maximal set of documents $\docs$ that is accessible by all entities in $\mathcal{A}$ using Algorithm \ref{alg:expd} (line 4). Then, it finds the maximal set of entities $\entities$ that can access all the documents in $\docs$ using Algorithm \ref{alg:expsg} (line 5). This process essentially generates a candidate maximal biclique $(\entities, \docs)$ from the initial seed $\mathcal{A}$. Candidate bicliques not satisfying the size thresholds $\tau_{\entities}, \tau_{\docs}$ are filtered out (line 6), and the pair that maximizes the product of the sizes is selected $(\entities_{res}, \docs_{res})$ and returned (lines 10-11).

The approach detailed in Algorithm \ref{alg:maximal} for finding a maximal biclique is particularly well-suited to common enterprise scenarios where there is an orders-of-magnitude difference between the number of documents in $\docsall$ and the number of entities in $\entall$, with a relatively small number of entity sets governing access to the large document corpus. In such a setting, it is likely that one of the unique ACLs $\mathcal{A}$, when used as an initial seed in Algorithm \ref{alg:maximal} (line 3), corresponds exactly to the entity set of the true maximum biclique. If this condition holds, then the biclique corresponding to the returned $(\entities_{res}, \docs_{res})$ in Algorithm \ref{alg:maximal} (line 11) is guaranteed to be the maximum biclique. 

\paragraph{Time complexity.}
The overall time complexity is roughly proportional to the number of unique ACLs multiplied by the number of documents and the logarithm of the maximum number of entities, $O(|\docsall|{^2}\log{|\entall|})$, due to nested iteration over all ACLs and set operations in Algorithm \ref{alg:expd} within the main loop.



\end{document}